# Interval Parsing Grammars for File Format Parsing

JIALUN ZHANG, Pennsylvania State University, USA
GREG MORRISETT, Cornell University, USA
GANG TAN, Pennsylvania State University, USA

File formats specify how data is encoded for persistent storage. They cannot be formalized as context-free grammars since their specifications include context-sensitive patterns such as the random access pattern and the type-length-value pattern. We propose a new grammar mechanism called Interval Parsing Grammars (IPGs) for file format specifications. An IPG attaches to every nonterminal/terminal an interval, which specifies the range of input the nonterminal/terminal consumes. By connecting intervals and attributes, the context-sensitive patterns in file formats can be well handled. In this paper, we formalize IPGs' syntax as well as its semantics, and its semantics naturally leads to a parser generator that generates a recursive-descent parser from an IPG. In general, IPGs are declarative, modular, and enable termination checking. We have used IPGs to specify a number of file formats including ZIP, ELF, GIF, PE, and part of PDF; we have also evaluated the performance of the generated parsers.

## 1 INTRODUCTION

A file format is a standard way to specify how data is encoded for persistent storage. Many file formats have been designed, e.g., GIF for images, ELF for binary executables, ZIP for archive files. To convert files in these formats to an in-memory form for further processing, we need to develop parsers based on format specifications. Most file format specifications are written in natural languages. Due to the imprecision and underspecification of natural languages, different interpretations may lead to semantic discrepancies between different parsers for the same file format. Such discrepancies can be exploited by attackers. For example, ZIP files that are prefixed by random garbage can still be extracted by `unzip` but fail to be recognized by a parser that conforms to the format specification [Jana and Shmatikov 2012].

Specifying file formats by formal grammars is a good first step towards the goal of eliminating the discrepancies. A formal grammar is a precise representation of a format specification, eliminating imprecision. Moreover, a parser can be automatically generated from a grammar, avoiding incorrect implementations. A successful example is context-free grammars (CFGs) in specifying the syntax of programming languages. Also, CFGs enable highly automated parser generation in parser frameworks such as YACC and ANTLR [Parr et al. 2014].

Authors' addresses: Jialun Zhang, Pennsylvania State University, USA, jjz5354@psu.edu; Greg Morrisett, Cornell University, USA, jgm19@cornell.edu; Gang Tan, Pennsylvania State University, USA, gtan@psu.edu.

However, file formats have context-sensitive patterns that cannot be described by CFGs. One notable pattern is the *random access pattern*, which requires a parser to jump to an offset given by a previously parsed number. For instance, in the ELF format, a header is at the beginning of an ELF file and an offset is stored in the header telling the start offset of the file's section header table. A parser needs to first parse the header, then use that offset to find the section header table. Another common pattern is the *type-length-value pattern*, which starts with a fixed-size type description and a fixed-size length, followed by a value field that should be parsed according to the type and with the length. These kinds of context sensitivity are essential to file formats and cannot be specified by CFGs.

To accommodate these common patterns that go beyond CFGs, data-dependent grammars [Fisher and Gruber 2005; Fisher et al. 2006; Jim et al. 2010; Mundkur et al. 2020; Ramananandro et al. 2019] and data-description languages [Back 2002; Bangert and Zeldovich 2014; Hmelnov and Mikhailov 2019; Kaitai 2015] allow certain forms of dependencies in file-format specifications. However, data-dependent grammars assume the left-to-right parsing order and are not expressive enough for random access, which can move the parsing position backward. Data-description languages handle random access in an imperative way, which leads to nontermination and modularity issues. A detailed analysis of their limitations can be found in the related-work section (section 6).

To specify the syntax and parsing-related semantics of file formats while addressing the shortcomings of previous research, we propose Interval Parsing Grammars (IPGs). An IPG has a similar syntax as a context-free grammar with attributes [Knuth 1968], but each nonterminal/terminal is annotated with an interval, which is a pair of expressions that evaluate to integers. The interval specifies the range of input that should be used for parsing according to the rule of the nonterminal/terminal. Moreover, attributes can be used in the intervals so that we can direct the parsing using previously parsed data. Using the random access pattern as an example, the offset in the header can be parsed and stored in an attribute and later used in an interval for the section header table. In general, an IPG can be used to automatically generate a recursive descent parser and each nonterminal is a subparser with input specified by its interval. The dependence between different parts of a file format can be expressed as the dependence between attributes and between attributes and intervals.

To the best of our knowledge, IPGs support all syntactic and parsing-based properties in common file formats and can reduce discrepancies between a file format specification and an implementation, as well as discrepancies between different implementations. However, we note that there are additional semantic properties that may be imposed by a file format. For instance, the SVG (Scalable Vector Graphics) format requires that objects in an SVG file should not have cycles via the `href` attributes of those objects [W3C 2018]. These semantic properties can be validated by a separate pass [Kumar et al. 2023] after parsing and is out of the scope of this paper.

In general, our paper makes the following contributions:

- We propose IPGs to specify file formats. We formalize the syntax and semantics of IPGs and implement a parser generator that generates a recursive descent parser from an IPG. We further propose a termination checking algorithm that checks if the parsing of an IPG terminates.
- For evaluation, we use IPGs to develop parsers for common file formats, and show that the context-sensitive patterns in file formats can be well handled. We further compare the performance of our parsers with hand-written parsers and parsers developed in related tools and show that our method achieves competitive performance.

The remaining sections are organized as follows. First, section 2 introduces features of IPGs using concrete examples. Second, section 3 formally defines the syntax and semantics of IPGs. After

$$S \rightarrow A[0,2]\ B[\mathbf{EOI}-2,\mathbf{EOI}]$$
$$A \rightarrow \texttt{"aa"}[0,2]$$
$$B \rightarrow \texttt{"bb"}[0,2]$$

Fig. 1. The first example.

introducing IPGs, we do a case study on specifying real file formats in section 4. Then we propose a termination checking algorithm for IPGs in section 5. In section 6, we compare existing works with IPGs and show how their problems are solved by IPGs. Finally, we evaluate IPGs in the parsing of different file formats in section 7.

## 2 IPG OVERVIEW

An Interval Parsing Grammar (IPG) is similar to a Context-Free Grammar (CFG) with a few design changes. The most important change is that intervals are assigned to nonterminals and terminal strings to specify the ranges of input they should describe. In addition, we borrow the idea of attributes from attribute grammars [Knuth 1968]. By using attribute values in intervals, IPGs enable the parsing to be dependent on previously parsed data. In this section, we use some toy examples to show the basic mechanisms of IPGs and how they specify some common patterns in file formats. Also, using these examples, we show the differences between IPGs and CFGs, and why those patterns cannot be described by CFGs.

*The First Example.* The toy example in Figure 1 shows how intervals work in IPGs. $S$ is the start nonterminal that receives the entire input. In the rule of $S$ (the first line), nonterminals $A$ and $B$ have assigned intervals. Term $A[0,2]$ means that the grammar rule of nonterminal $A$ should describe a slice of the input string received by $S$ from index 0 to index 2 (with 2 not included). Symbol **EOI** is End-Of-Input, i.e., the length of the input received by the current nonterminal (the one on the left-hand side of the rule). Therefore, $B[\mathbf{EOI}-2,\mathbf{EOI}]$ means that the rule of $B$ should describe the last two characters of the input to $S$. The rule of $A$ checks if the first two characters match the terminal string `"aa"`. The rule of $B$ is similar. In summary, this grammar accepts any string in the form of `"aa...bb"`.

This example shows one difference between IPGs and CFGs. In CFGs, when there is a concatenation of two nonterminals $AB$, the input described by $B$ starts right after where $A$ finishes. In contrast, IPGs specify the input range by assigning intervals to nonterminals, allowing nonterminals to describe strings that are specific slices of the input.

*Random Access Pattern.* In the last example, **EOI** is introduced to specify a position relative to the end of the input. However, this is often not enough in file formats. One prevalent pattern in file formats is the *random access pattern*, which includes an offset somewhere in the file and directs the parser to move the parsing position to that offset. As an example, The Executable and Linkable Format [ELF 1995] has a fixed-size header at the beginning, which contains an offset pointing to the section table. The section table then has an array of offsets pointing to the beginnings of sections. Following these offsets, an ELF parser can find all the sections. This process involves random accesses that direct the parser to jump to different parts of the input file. To handle random accesses, IPGs adopt the idea of attributes and allow the use of attribute values in intervals.

For illustration, suppose a toy file format starts with an 8-byte header. Inside the header, there are two 4-byte numbers that tell the starting offset and the size of some data in the file. This format can be specified by the IPG in Figure 2. For simplicity, we omit the rules for *Int* and *Data* and assume that *Int* specifies an integer and the value of the integer is stored in the attribute $Int.val$. The IPG

$$
\begin{aligned}
S &\longrightarrow H[0,8]\ \mathit{Data}[H.\mathit{offset}, H.\mathit{offset} + H.\mathit{length}] \\
H &\longrightarrow \mathit{Int}[0,4]\ \{\mathit{offset}{=}\mathit{Int}.\mathit{val}\} \\
&\qquad \mathit{Int}[4,8]\ \{\mathit{length}{=}\mathit{Int}.\mathit{val}\}
\end{aligned}
$$

Fig. 2. Random access pattern.

$$
\begin{aligned}
\mathit{Int} &\longrightarrow \mathit{Int}[0, \mathbf{EOI}-1]\ \mathit{Digit}[\mathbf{EOI}-1, \mathbf{EOI}] \\
&\qquad \{\mathit{val}{=}2 * \mathit{Int}.\mathit{val} + \mathit{Digit}.\mathit{val}\} \\
&\quad /\ \ \mathit{Digit}[0,1]\ \{\mathit{val}{=}\mathit{Digit}.\mathit{val}\} \\
\mathit{Digit} &\longrightarrow \texttt{"0"}[0,1]\ \{\mathit{val}{=}0\} \\
&\quad /\ \ \texttt{"1"}[0,1]\ \{\mathit{val}{=}1\}
\end{aligned}
$$

Fig. 3. Binary number parser.

rule for $S$ in Figure 2 uses $H[0, 8]$ on the first 8 bytes of the input for describing the header. In the rule for $H$, the first integer is stored as attribute $H.\mathit{offset}$ and the second as attribute $H.\mathit{length}$. Back to the rule of $S$, the information in the header is used to locate *Data*, which starts at $H.\mathit{offset}$ and ends at $H.\mathit{offset} + H.\mathit{length}$.

This example shows how to store parsing results as attributes and use attribute values in intervals. In this way, the previously parsed data can direct the following parsing, which is context-sensitive and beyond the ability of CFGs. Even data-dependent grammars [Jim et al. 2010] cannot express the random access pattern because they cannot express dependencies between control (the parsing position) and data.

*Parser for a Binary Number.* Figure 3 presents an IPG for describing *Int*s used in the last example; it accepts a string representing a binary number and also computes the attribute $\mathit{Int}.\mathit{val}$ as the value of that binary number. Note that alternatives in IPGs are separated by slashes, which are *biased choices* as in PEGs [Ford 2004a]. If an earlier alternative succeeds, later alternatives will not be tried; as a result, ambiguity is eliminated. In this example, when *Int* receives the entire input, it tries the first alternative. In this alternative, it recursively tries *Int* on the entire input except for the last byte, indicated by the interval $[0, \mathbf{EOI} - 1]$. The recursion ends when the length of the local input is 0 and the interval computation fails since $[0, \mathbf{EOI} - 1] = [0, -1]$—an invalid interval [1]. Then the second alternative of *Int* is also tried and failed because the interval $[0, 1]$ exceeds the range of the input. In this case, the parser falls back to the previous level with the input interval $[0, 1]$ and tries the second alternative of *Int* , which accepts only one character by *Digit*. The attributes in this example are used to compute the integer values; their definitions are straightforward.

The example shows how to specify a repetitive pattern by recursion in IPGs. It shares a similar pattern as CFGs. Actually, if we remove intervals and attributes and change the biased choice to the unbiased choice, we get a CFG defining the language $(0|1)^+$. However, due to left recursion in the resulting CFG, a simple recursive descent parser would not work for this CFG. But it is not the case for the IPG, because each time *Int* is invoked during parsing, its given input is shorter and the left recursion stops when the length of the input is 1. With a similar kind of reasoning, we can statically prove the parsing termination of many IPGs.

[1] $[0, 0]$ is still valid because we need it to represent empty intervals for the empty string.

$$
\begin{aligned}
S &\rightarrow \texttt{"1"}\texttt{[}0,1\texttt{]}\ O\texttt{[}1,\mathbf{EOI}\texttt{]}\ \texttt{"stop"}\texttt{[}O.\mathbf{end},\mathbf{EOI}\texttt{]} \\
O &\rightarrow \texttt{"0"}\texttt{[}0,1\texttt{]}\ O\texttt{[}1,\mathbf{EOI}\texttt{]} \\
&\ \ /\ \texttt{"0"}\texttt{[}0,1\texttt{]}
\end{aligned}
$$

Fig. 4. Example showing special attributes.

$$
\begin{array}{lll}
\text{Grammar} & G & ::= R_1 \ldots R_n \\
\text{Rule} & R & ::= A \rightarrow alt_1 / \ldots / alt_n \\
\text{Alternative} & alt & ::= tm_1 \ldots tm_n \\
\text{Term} & tm & ::= A\texttt{[}e_l, e_r\texttt{]} \ \mid\ s\texttt{[}e_l, e_r\texttt{]} \ \mid\ \texttt{\{}id\texttt{=}e\texttt{\}} \ \mid\ \langle e \rangle \\
 & & \quad \mid\ \texttt{for}\ id\texttt{=}e_1\ \texttt{to}\ e_2\ \texttt{do}\ A\texttt{[}e_l, e_r\texttt{]} \\
 & & \\
\text{Expr} & e & ::= n \mid e_1\ bop\ e_2 \mid e_1 ? e_2 : e_3 \mid ref \\
\text{Binary Op} & bop & ::= + \mid - \mid * \mid / \mid = \mid > \mid < \mid \land \mid \lor \\
\text{Reference} & ref & ::= id \mid A.id \mid A(e).id \mid \mathbf{EOI} \mid A.\mathbf{start} \mid A.\mathbf{end}
\end{array}
$$

Fig. 5. Core IPG language syntax; $A$ and $B$ for nonterminals; $s$ for terminal strings; $id$ for attributes.

*Special Attributes.* As we have shown, IPGs add an interval for each nonterminal and terminal string. However, sometimes we want the input of a nonterminal $B$ starts right after where the parsing of $A$ ends, with the parsing of $A$ consuming an arbitrary-length input. Such behavior can be well captured by a CFG rule like $S \rightarrow AB$. To allow this flexibility, we introduce a special attribute **end** to emulate this CFG-like parsing behavior. Intuitively, $A.\mathbf{end}$ means one plus the offset of the right-most character touched by the parsing of $A$. For example, Figure 4 specifies a parser that feeds $O$ with the entire input except for the first character and the parsing of $O$ touches only a prefix of the interval $[1, \mathbf{EOI}]$. After input is consumed by $O$, a terminal `"stop"` immediately follows. Consequently, the parser accepts any string in the form of `"10...0stop"`. In the same spirit, **start** is defined as the offset of the left-most character touched by the parsing of a nonterminal. We formally describe the computation of these two special attributes in section 3.

To summarize, with attributes and intervals, IPGs allow the user to control how the parsing position moves and express dependencies between different parts of the input. These abilities are required by many file formats but are out of the capability of CFGs.

## 3 IPG SYNTAX AND SEMANTICS

We start by formalizing the syntax and semantics of a core IPG language, before briefly discussing other features such as auto-completion of intervals in the full IPG language.

### 3.1 Core Language Syntax

The syntax of the core IPG language is defined in Figure 5, where we use metasymbols $A$ and $B$ for nonterminals, $id$ for attribute names, and $s$ for terminal strings. A grammar $G$ consists of one or more rules. Each rule $R$ has a nonterminal on the left-hand side and an ordered list of alternatives separated by slashes (biased choice). We assume there is exactly one rule for each nonterminal.

Each alternative in a rule is a list of terms. The first three kinds of terms (nonterminals with intervals, terminal strings with intervals, and attribute definitions) have been discussed in section 2. Note that a terminal string can be an empty string ($\epsilon$). In addition, $\langle e \rangle$ represents a predicate with a

```
1  S → H[0, 4] {size=4}
2        for i=0 to H.num do A[4 + size ∗ i, 4 + size ∗ (i + 1)]
3        {a0=A(0).val}
4        ⟨ a0>0 ∧ a0<10 ⟩
5  H → Int[0, 4] {num=Int.val}
6  A → Int[0, 4] {val=Int.val}
```

Fig. 6. An IPG example.

boolean formula $e$, which fails if $e$ evaluates to false. An example is at Line 4 in Figure 6; it checks if the attribute $a_0$ falls between 0 and 10. The last kind "**for** $id$=$e_1$ **to** $e_2$ **do** $A[e_l, e_r]$" is an array term, which accepts a sequence of array elements, with each element specified by $A[e_l, e_r]$; expressions $e_l$ and $e_r$ may contain occurrences of loop variable $id$, which ranges from $e_1$ to $e_2 - 1$. Note that when $e_2$'s value is less than or equal to $e_1$'s value, the for-loop for an array term does not run; it imposes no constraints and accepts any string. Line 2 in Figure 6 gives an example of an array, which repeats $A$ for $H.num$ times; the length of each repeat is of length *size*.

An expression is for expressing the computation of attribute values and intervals. An expression can be a natural number $n$, a binary operation over two expressions, a ternary conditional expression ($e_1 ? e_2 : e_3$), or an attribute reference. An attribute reference *ref* is a reference to the value of an attribute. There are six forms. When used without prefixing a nonterminal (the first case), it refers to an attribute in the same alternative of a rule (e.g., *size* at line 2 in Figure 6), or a loop variable in the current context (e.g., $i$ at line 2 in Figure 6). The second case $A.id$ refers to the attribute $id$ of a nonterminal $A$ in the same alternative of a rule. At line 2 in Figure 6, $H.num$ provides an example of this case. The third case ($A(e).id$) refers to the attribute of an array element, which is used at line 3 in Figure 6 to refer to an attribute of the first element of the array. The rest three cases are special attributes End-Of-Input (**EOI**), **start**, and **end**. Their usage has been shown in section 2. All expressions in our language evaluate to integers. It would be a straightforward extension to add more types of expressions and introduce a type system. For example, boolean attributes could be introduced to be used as the conditions of if-then-else expressions.

### 3.2 Attribute Checking

IPG attribute checking ensures two properties: (1) every attribute reference refers to a properly defined attribute, and (2) there are no circular definitions among attributes.

The first property is checked straightforwardly in two steps. In the first step, for a nonterminal $A$, the attribute checker collects a set of defined attributes for $A$; we write $def(A)$ for the set. This is performed by collecting what attributes are defined in the IPG rule for $A$. Recall that the rule may have multiple alternatives; $def(A)$ is the set of attributes that are defined in all alternatives. In the second step, the checker checks that every attribute reference in the grammar refers to a defined attribute. Particularly, if the reference is of the form $B.id$ or $B(e).id$, it ensures $id \in def(B)$; if the reference is of the form $id$ and it appears in the rule for nonterminal $A$, it ensures $id \in def(A)$.

For the second property of no circularity in attribute definitions, let us first inspect an example. Suppose a rule for a nonterminal $A$ has an alternative "$B_1[0, B_2.a]$ $B_2[a_1, \textbf{EOI}]$ $\{a_1{=}2\}$". It introduces for $A$ an attribute $a_1$, used as the left interval of $B_2$, and defines $B_1$'s right interval as the attribute value of $a$ for $B_2$; here we assume that the rule for $B_2$ (not shown) defines an attribute $a$. This flexibility of what attributes can be used in attribute and interval definitions allows patterns such as backward parsing. However, this can lead to circular dependencies. To avoid that, attribute checking builds a dependency graph for each alternative and rejects IPGs that have an alternative whose dependency graph is not a DAG (directed acyclic graph).

In detail, for an alternative, we define its dependency graph as $G = (V, E)$, where $V$ is the set of terms in this alternative. When a term $t_1$ contains a reference to an attribute of another term $t_2$ in the alternative, we add an edge $t_1 \rightarrow t_2$ into the graph. For the previous example, the dependency graph is $G = (\{B_1, B_2, a_1\}, \{B_2 \rightarrow a_1, B_1 \rightarrow B_2\})$. Since it is a DAG, the dependencies are acceptable. Given the dependency graph of an alternative, our parser generator further reorders terms in the alternative by the dependency graph's topological order. For the example, the reordered alternative is "$\{a_1{=}2\}$ $B_2[a_1, \textbf{EOI}]$ $B_1[0, B_2.a]$". In the following discussion about parsing semantics, we assume all alternatives have already been reordered by the aforementioned procedure.

### 3.3 Core Language Parsing Semantics

In this section, we formalize the parsing semantics of IPGs. The semantics is similar to recursive descent parsing for a CFG, but with a couple of important differences. First, IPGs use biased choice, which means that alternatives in a rule are tried sequentially and the parsing stops at the first successful alternative. Second, for terminals and nonterminals, their assigned intervals are computed first and subparsers can inspect only the local input confined by those intervals.

For notation, we write $\overrightarrow{m}$ as a sequence of $m$-s, where $m$ can be an arbitrary entity (e.g., terms). We write $\varepsilon$ as the empty sequence and overload "$\cdot$" for both adding an element to a sequence ($m_1 \cdot \overrightarrow{m}$) and sequence concatenation ($\overrightarrow{m_1} \cdot \overrightarrow{m_2}$). When formalizing the parsing semantics of IPGs, we treat the right-hand side of a rule as a sequence of alternatives, written as $\overrightarrow{alt}$, and each alternative as a sequence of terms, written as $\overrightarrow{tm}$. We also use $E$ for an *environment* that maps from attribute IDs to values. We write $\{id \mapsto v\}$ for the environment only with the map from $id$ to $v$ and $E[id \mapsto v]$ for the environment that adds to $E$ the map from $id$ to $v$.

Parsing an input string according to an IPG grammar produces either Fail or a parse tree, defined as follows:

$$\text{Parse tree} \quad Tr ::= \text{Node}(A, E, \overrightarrow{Tr}) \mid \text{Array}(\overrightarrow{Tr}) \mid \text{Leaf}(s)$$

Tree "$\text{Node}(A, E, \overrightarrow{Tr})$" has nonterminal $A$ as the root and $\overrightarrow{Tr}$ as the sequence of child parse trees, and $E$ is an *environment* that records attribute values for $A$'s attributes. Tree $\text{Array}(\overrightarrow{Tr})$ represents the result of parsing an array and the parse trees of array elements are $\overrightarrow{Tr}$. Finally, $\text{Leaf}(s)$ is the parse tree for matching a terminal string $s$.

We formalize IPG parsing semantics as a big-step relation, with judgments listed in Figure 7 and rules listed in Figure 8. The rule for "$s \vdash A \Downarrow R$" looks up $A$'s rule in input grammar $G$ and then performs parsing according to $A$'s alternatives $\overrightarrow{alt}$. The rules for "$s, A \vdash \overrightarrow{alt} \Downarrow R$" implement the biased choice semantics: it checks if parsing according to the first alternative succeeds; if so, it succeeds with the result of the first alternative; otherwise, it tries the rest.

The rules for "$s, A, E, \overrightarrow{Tr} \vdash \overrightarrow{tm} \Downarrow R$" are more complex because terms that appear later may use attributes defined in earlier terms. Therefore, it is necessary to pass the current environment for $A$ and parse trees $\overrightarrow{Tr}$ of earlier terms as parameters. For example, the rule A-Seq1 passes the current environment $E$ to the first term $tm_1$, then passes the updated term $E_1$ and the resulting parse trees $\overrightarrow{Tr} \cdot Tr_1$ to the rest of the terms.

The rules for "$s, A, E, \overrightarrow{Tr} \vdash tm \Downarrow E', R_{\text{tm}}$" processes individual terms. In these rules, we write $\sigma(E, \overrightarrow{Tr}, e)$ for the result of evaluating $e$ under environment $E$ and parse trees $\overrightarrow{Tr}$; note $e$ can use attributes in $E$ as well as those that appear in the environments of the root nodes of $\overrightarrow{Tr}$. Most rules are self explanatory and we explain only a couple next.

| | |
|---|---|
| $s \vdash A \Downarrow R$ | parsing input $s$ according to nonterminal $A$'s rule produces result $R$. |
| $s, A \vdash \overrightarrow{alt} \Downarrow R$ | parsing input $s$ according to alternatives $\overrightarrow{alt}$ produces result $R$, assuming $\overrightarrow{alt}$ is part of $A$'s rule. |
| $s, A, E, \overrightarrow{Tr} \vdash \overrightarrow{tm} \Downarrow R$ | parsing input $s$ according to a sequence of terms $\overrightarrow{tm}$ produces result $R$, assuming $\overrightarrow{tm}$ belong to an alternative of $A$ and can use attributes in $E$ and $\overrightarrow{Tr}$. |
| $s, A, E, \overrightarrow{Tr} \vdash tm \Downarrow E', R_{\text{tm}}$ | similar to the last one, except for a single term and the result includes the updated environment $E'$. |

Fig. 7. Judgments for IPG parsing semantics.

The rule T-Ter matches a terminal string $s_1$ with interval $[e_l, e_r]$. It first evaluates $e_l$ and $e_r$ to get the interval $[l, r]$, checks if the interval falls in the range of the input, and, if it is, checks if the input string from $l$ to $l + |s_1| - 1$ matches $s_1$; if either of the checks fails, failure is raised. Furthermore, since parsing by this rule touches the input from $l$ to $l + |s_1| - 1$ (or touches nothing if $s_1 = \epsilon$), it updates the environment's **start** and **end** attributes (or keeps the same environment if $s_1 = \epsilon$) with the following definition:

$$\text{updStartEnd}(E, l, r, b) = \begin{cases} E[\textbf{start} \mapsto min(E[\textbf{start}], l), \textbf{end} \mapsto max(E[\textbf{end}], r)], & \text{if } b \text{ holds;} \\ E, & \text{otherwise.} \end{cases}$$

The rule T-NTSucc matches a nonterminal $B$ with an interval. It is similar to T-Ter, except for a few differences. First, the substring of input from $l$ to $r - 1$ is parsed by the grammar rule of $B$. Second, $B$'s **start** and **end** are adjusted by adding $l$, to switch from relative offsets within $s[l, r]$ to offsets within $s$. Using Figure 4 as a example, given input `"1000stop"`, in the first rule, after the parsing of $O[1, \textbf{EOI}]$, $O.\textbf{end} = 3$; it needs to be adjusted to add 1 to become 4. Therefore the parsing of `"stop"`$[O.\textbf{end}, \textbf{EOI}]$ starts at the correct offset 4. Finally, to deal with the corner case when the parsing of $B$ does not touch any input, the environment is updated only if $B$'s **end** value is not zero; this is formalized with the updStartEnd as in the T-Ter case. Rules for Array($\overrightarrow{Tr}$) are straightforward and are left to the appendix in this paper's Arxiv version [?].

The parsing semantics in Figure 8 is formalized as an interpreter for IPGs. Since the rules are syntax directed according to IPG syntax, we can reformulate it as a parser generator, which is how our IPG parser generator is implemented. Essentially, we define a generator for each syntactic category: GenNT is the generator for nonterminals, GenAlts is the generator for a sequence of alternatives, etc. Then the rules in Figure 8 can be converted to generator definitions. E.g., $\text{GenNT}(A) = \lambda s.\ \text{GenAlts}(\overrightarrow{alt}, A, s)$, if $A = \overrightarrow{alt}$ is the rule for $A$. This process is straightforward and we omit the details. Furthermore, adopting the memoization idea from PEG parsing [Ford 2004a], the resulting parser's time complexity is $O(n^2)$ for the core IPG language, where $n$ is the length of the input. The generated parser remembers the parsing result of $S[l, r]$ (i.e., a slice of input $S$ from $l$ to $r$) with some start nonterminal; in this way, next time when the parser tries to parse $S[l, r]$ with the same start nonterminal, the result can be retrieved from the memoization table. As a result, assuming the grammar size is constant, the generated parser's time complexity is $O(n^2)$.

To further evaluate the idea of specifying intervals in parsing, we also implemented a parser combinator library based on this idea in OCaml. The detail of this library can be found in the appendix of the Arxiv version [?].

$\boxed{s \vdash A \Downarrow R}$

$$\frac{(A \rightarrow \overrightarrow{alt}) \in \text{rules}(G) \quad s, A \vdash \overrightarrow{alt} \Downarrow R}{s \vdash A \Downarrow R} \text{ (G-NT)}$$

$\boxed{s, A \vdash \overrightarrow{alt} \Downarrow R}$

$$\frac{s, A, \{\textbf{EOI} \mapsto |s|, \textbf{start} \mapsto |s|, \textbf{end} \mapsto 0\}, \varepsilon \vdash alt_1 \Downarrow Tr}{s, A \vdash alt_1 \cdot \overrightarrow{alt} \Downarrow Tr} \text{ (R-AltSucc)} \qquad \frac{}{s, A \vdash \varepsilon \Downarrow \text{Fail}} \text{ (R-Emp)}$$

$$\frac{s, A, \{\textbf{EOI} \mapsto |s|, \textbf{start} \mapsto |s|, \textbf{end} \mapsto 0\}, \varepsilon \vdash alt_1 \Downarrow \text{Fail} \quad s, A \vdash \overrightarrow{alt} \Downarrow R}{s, A \vdash alt_1 \cdot \overrightarrow{alt} \Downarrow R} \text{ (R-AltFail)}$$

$\boxed{s, A, E, \overrightarrow{Tr} \vdash \overrightarrow{tm} \Downarrow R}$

$$\frac{\begin{array}{c} s, A, E, \overrightarrow{Tr} \vdash tm_1 \Downarrow E_1, Tr_1 \\ s, A, E_1, \overrightarrow{Tr} \cdot Tr_1 \vdash \overrightarrow{tm} \Downarrow R \end{array}}{s, A, E, \overrightarrow{Tr} \vdash tm_1 \cdot \overrightarrow{tm} \Downarrow R} \text{ (A-Seq1)} \qquad \frac{\begin{array}{c} s, A, E, \overrightarrow{Tr} \vdash tm_1 \Downarrow E_1, \varepsilon \\ s, A, E_1, \overrightarrow{Tr} \vdash \overrightarrow{tm} \Downarrow R \end{array}}{s, A, E, \overrightarrow{Tr} \vdash tm_1 \cdot \overrightarrow{tm} \Downarrow R} \text{ (A-Seq2)}$$

$$\frac{}{s, A, E, \overrightarrow{Tr} \vdash \varepsilon \Downarrow \text{Node}(A, E, \overrightarrow{Tr})} \text{ (A-Emp)} \qquad \frac{s, A, E, \overrightarrow{Tr} \vdash tm_1 \Downarrow E, \text{Fail}}{s, A, E, \overrightarrow{Tr} \vdash tm_1 \cdot \overrightarrow{tm} \Downarrow \text{Fail}} \text{ (A-Fail)}$$

$\boxed{s, A, E, \overrightarrow{Tr} \vdash tm \Downarrow E', R_{\text{tm}}}$

$$\frac{\begin{array}{c} \sigma(E, \overrightarrow{Tr}, e_l) = l \qquad \sigma(E, \overrightarrow{Tr}, e_r) = r \\ 0 \le l \le r \le |s| \qquad r - l \ge |s_1| \qquad s[l, l + |s_1|] = s_1 \end{array}}{s, A, E, \overrightarrow{Tr} \vdash s_1[e_l, e_r] \Downarrow \text{updStartEnd}(E, l, r, s_1 \neq \epsilon), \text{Leaf}(s_1)} \text{ (T-Ter)}$$

$$\frac{\sigma(E, \overrightarrow{Tr}, e_l) = l \qquad \sigma(E, \overrightarrow{Tr}, e_r) = r \qquad 0 \le l \le r \le |s| \qquad s[l, r] \vdash B \Downarrow \text{Node}(B, E_B, \overrightarrow{Tr_B})}{\begin{array}{l} s, A, E, \overrightarrow{Tr} \vdash B[e_l, e_r] \Downarrow \text{updStartEnd}(E, l + E_B[\textbf{start}], l + E_B[\textbf{end}], E_B[\textbf{end}] \neq 0), \\ \qquad \text{Node}(B, E_B[\textbf{start} \mapsto l + E_B[\textbf{start}], \textbf{end} \mapsto l + E_B[\textbf{end}]], \overrightarrow{Tr_B}) \end{array}} \text{ (T-NTSucc)}$$

$$\frac{\sigma(E, \overrightarrow{Tr}, e) = v}{s, A, E, \overrightarrow{Tr} \vdash \{id{=}e\} \Downarrow E[id \mapsto v], \varepsilon} \text{ (T-Attr)} \qquad \frac{\sigma(E, \overrightarrow{Tr}, e) \neq 0}{s, A, E, \overrightarrow{Tr} \vdash \langle e \rangle \Downarrow E, \varepsilon} \text{ (T-Pred)}$$

$$\frac{\sigma(E, \overrightarrow{Tr}, e_l) = l \qquad \sigma(E, \overrightarrow{Tr}, e_r) = r \qquad \neg(0 \le l < r \le |s|) \ \vee \ s[l, r] \vdash B \Downarrow \text{Fail}}{s, A, E, \overrightarrow{Tr} \vdash B[e_l, e_r] \Downarrow E, \text{Fail}} \text{ (T-NTFail)}$$

$$\frac{\begin{array}{c} \sigma(E, \overrightarrow{Tr}, e_l) = l \qquad \sigma(E, \overrightarrow{Tr}, e_r) = r \\ \neg(0 \le l \le r \le |s|) \ \vee \ r - l < |s_1| \\ \vee \ s[l, l + |s_1|] \neq s_1 \end{array}}{s, A, E, \overrightarrow{Tr} \vdash s_1[e_l, e_r] \Downarrow E, \text{Fail}} \text{ (T-TerFail)} \qquad \frac{\sigma(E, \overrightarrow{Tr}, e) = 0}{s, A, E, \overrightarrow{Tr} \vdash \langle e \rangle \Downarrow E, \text{Fail}} \text{ (T-PredFail)}$$

Fig. 8. IPG parsing semantics.

### 3.4 Full IPG Language

We describe additional features we implement in the full IPG language; they are useful for describing various file formats.

*Implicit Intervals.* Writing IPGs becomes tedious when we must specify intervals for every nonterminal and terminal string. With the special attribute **end**, we can simplify the writing of IPGs by omitting some obvious intervals. We have implemented an auto-completion feature that fills out missing intervals. For example,

$$S \rightarrow \texttt{"magic"}\ A\ B\texttt{[}10\texttt{]}$$

can be completed as

$$S \rightarrow \texttt{"magic"}\texttt{[}0,5\texttt{]}\ A\texttt{[}5,\mathbf{EOI}\texttt{]}\ B\texttt{[}A.\mathbf{end}, A.\mathbf{end}+10\texttt{]}.$$

For an alternative with a sequence of terms, auto completion scans from left to right and infers the missing intervals of a term based on its last term. For a nonterminal, if both endpoints of its interval are missing, the left endpoint is inferred to be the **end** of the last term if it is a nonterminal (e.g., $A$.**end** as $B$'s left endpoint), or the right endpoint of the last term if it is a terminal string (e.g., 5 as $A$'s left endpoint). And the right endpoint is **EOI** (e.g., **EOI** as $A$'s right endpoint). If there is only one expression between the parentheses, it is viewed as the length of the interval; the inference of the left endpoint is the same as the previous case, but the right endpoint is completed as the left endpoint plus the given length (e.g., $A$.**end** + 10). For a terminal string, since its length is known, we need to infer only the left endpoint, which is treated the same as the nonterminal case; its right endpoint will be the left endpoint plus the length of the string. One special case is for the left-most term, its left endpoint is inferred as 0 if it is missing.

*Local Rules.* Recall that one constraint of IPGs is that an attribute reference in an alternative of a rule can refer to only those attributes defined in the alternative. This constraint can sometimes result in grammar rules that are not easy to read. For convenience, we introduce **`where`** clauses to let users define local rules. As a toy example,

$$S \rightarrow A\ B\texttt{[}A.val, \mathbf{EOI}\texttt{]}\ C\texttt{[}B.val, \mathbf{EOI}\texttt{]}$$

can be rewritten as

$$S \rightarrow A\ D\texttt{[}0, \mathbf{EOI}\texttt{]}\ \texttt{where}\ D \rightarrow B\texttt{[}A.val, \mathbf{EOI}\texttt{]}\ C\texttt{[}B.val, \mathbf{EOI}\texttt{]}.$$

Note that we cannot simply create another rule for $D$ because $B$ depends on $A.val$. But inside the local rule of $D$ defined after **`where`**, $A.val$ is still visible. With local rules, one complicated alternative can be written in a more readable way. Readers may refer to the ELF format in section 4 for a more practical example.

*Switch Terms.* The *type-length-value pattern*, common in file formats, has a type description and a length, followed by data of that type and length. As an example, the EtherType field in an Ethernet frame tells the length of the following data (when $\leq 1500$) or represents the type of the protocol in use (when $\geq 1536$). Different subparsers should be chosen depending on the type information.

To specify such a behavior, we introduce switch terms. The syntax of a switch term is

$$\texttt{switch}(e_1 : A_1\texttt{[}e_{l_1}, e_{r_1}\texttt{]}\ /\ \ldots\ /\ e_n : A_n\texttt{[}e_{l_n}, e_{r_n}\texttt{]}\ /\ A_{n+1}\texttt{[}e_{l_{n+1}}, e_{r_{n+1}}\texttt{]}).$$

It has $n+1$ choices separated by slashes. Each of the first $n$ choices has a condition and a nonterminal with an interval. The semantics is similar to a switch statement in C. The choices are executed from left to right; if one of the conditions succeeds, then the corresponding nonterminal is used as the data description and the remaining choices are skipped. If all conditions fail, the default choice is used. Back to the EtherType example, with a switch term, a possible implementation can be

$$\texttt{switch}(\mathit{ethertype} \leq 1500 : \mathit{EtherType}[\mathit{ethertype}] \;/\; \mathit{ethertype} \geq 1536 : \ldots \;/\; \mathit{Fail}[1,0]),$$

which either uses *ethertype* as a length, or chooses different subparsers based on the value of *ethertype*. The default branch must fail because of its always-invalid interval. Note that switch terms are a syntactic sugar and can be implemented with a combination of predicates and biased choices.

*Existentials.* Sometimes it is useful to refer to an attribute of a particular element in an array, while only knowing that element satisfies some constraint. We introduce existentials $\exists id.e_1?e_2:e_3$ for that, where *id* is a loop variable identifier, $e_1$ is a condition on an array reference, and $e_2$ and $e_3$ are two expressions. Operationally, this expression loops over the referred array and stops at the first element that evaluates $e_1$ to true. If it finds one, loop variable *id* is set to that element's index and the evaluation goes to $e_2$, which may use *id*; if no element satisfying $e_1$ is found, the evaluation evaluates $e_3$. For example, suppose there is an array **for** $i$=0 **to** 10 **do** $Num[i, i+1]$; also *Num* has an attribute *val*, $Num(0).val = 1$ and $Num(1).val = 0$. Then $\exists j.Num(j) = 0?j:0$ evaluates to 1.

*Blackbox Parsers.* Another nice property provided by IPGs is modularity, which means legacy parsers can be reused as black boxes by passing them local input buffers specified by intervals. For example, in a GIF parser, once it determines where the compressed image data is, it can invoke an existing decompression algorithm to decode the code by passing it the input within the interval of the code area. By using an interval, the parser can control what can be seen by an external parser. A concrete example of blackbox parsers will be shown in section 7, which implements a decompressor for ZIP formats by calling decompression functions from zlib.

### 3.5 Relation between IPGs and CFGs

First of all, IPGs $\not\subseteq$ CFGs because the language $\{a^n b^n c^n | n > 0\}$ is not in CFGs but in IPGs. It can be specified by the following grammar.

$$\begin{aligned}
S &\longrightarrow \langle \mathbf{EOI} \bmod 3 = 0\rangle\ \{n{=}\mathbf{EOI}/3\}\ A[0,n]\ B[n,2n]\ C[2n,3n] \\
A &\longrightarrow \texttt{"a"}[0,1]\ A[1,\mathbf{EOI}] \;/\; \texttt{"a"}[0,1] \\
B &\longrightarrow \texttt{"b"}[0,1]\ B[1,\mathbf{EOI}] \;/\; \texttt{"b"}[0,1] \\
C &\longrightarrow \texttt{"c"}[0,1]\ C[1,\mathbf{EOI}] \;/\; \texttt{"c"}[0,1]
\end{aligned}$$

The relation between CFGs and IPGs is open because of IPGs' use of biased choice; this is similar to the open problem of the relationship between CFGs and PEGs. However, if we replaced biased choice in IPGs with unbiased choice, we could prove that CFGs $\subseteq$ IPGs. Intuitively, given a CFG, we can perform the implicit interval completion mentioned in section 3.4 to infer intervals for it and get an equivalent IPG.

## 4 CASE STUDIES

To show how IPGs can be used in practice, we present IPG specifications of some file formats in this section. File formats generally are in two main categories: directory-based and chunk-based. *Directory-based* formats resemble file systems and involve a large degree of random access. They often have a fixed-size header at a fixed location such as the start or the end of the file. The header contains an offset pointing to a table, which contains an array of offsets pointing to different pieces of data. Starting from the header, the parser can collect all data following those offsets. As examples, ELF, ZIP and OLE fall into this category. A file in a *chunk-based* format consists of a list of data chunks, and each chunk has a fixed-size header telling the type and length of it; i.e., a chunk follows

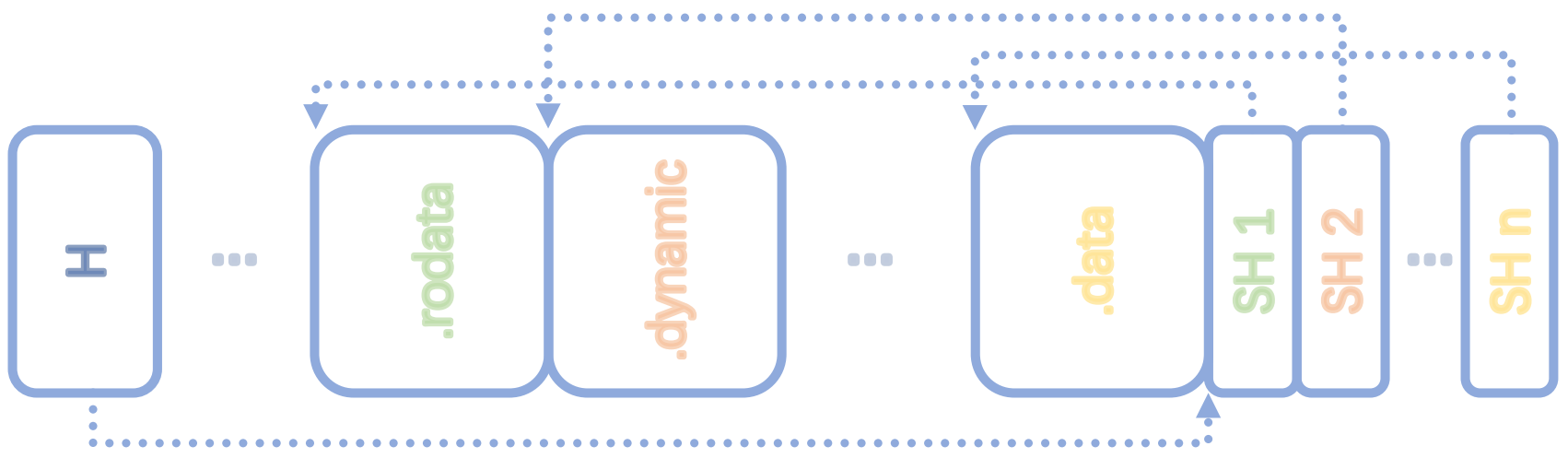

(a)

```
1  ELF    → H[128]
2           for i=0 to H.num do SH[H.ofs + i ∗ H.sz, H.ofs + (i + 1) ∗ H.sz]
3           for i=1 to H.num do Sec[SH(i).ofs, SH[i].ofs + SH[i].sz]
4           where Sec → switch(SH[i].type = 6 : DynSec / OtherSec)
5  H      → "0x7fELF" ...
6           Int[40, 48] {ofs=Int.val} ...
7           Int[58, 60] {sz=Int.val}
8           Int[60, 62] {num=Int.val} ...
9  SH     → ... Int[24, 32] {ofs=Int.val}
10          Int[32, 40] {sz=Int.val} ...
11 DynSec → for i=0 to EOI/16 do DynSecEntry[16 ∗ i, 16 ∗ (i + 1)]
```

(b)

Fig. 9. (a) The structure of an ELF file (arrows indicate offset pointers; H and SH are abbreviations for Header and Section Header, respectively). (b) Snippets of an IPG for the ELF format.

the type-length-value pattern. With the type information, the parser chooses the corresponding way to consume the following data of the specified length. After one data chunk is parsed, the same process repeats for the following chunks until the end of the file. Typically image formats adopt this design, including PNG, JPG and GIF.

In this section, we use ELF, GIF and PDF as examples. ELF and GIF are representatives of directory-based and chunk-based formats, respectively. PDF is picked because it is the most complicated format to our knowledge, which requires some unusual parser behaviors. We did not implement a full PDF parser due to its complexity, but a functional subset to show how IPGs can support some interesting features.

## 4.1 ELF

The Executable and Linkable Format [ELF 1995] is a format for executable files and object code and is prevalent on Unix-like operating systems such as Linux and OpenBSD, and mobile operating systems like Android. The general structure of ELF files is shown in Figure 9a. There are two views of an ELF file: the program view and the section view. Here we focus the discussion on the section view since the program view can be dealt with similarly.

In the section view, an ELF file starts with a fixed-size header, ends with a section header table, and contains many sections in between. The top-level rule starting at line 1 in Figure 9b shows these three structures. It starts with the header $H$, which contains the number of section headers ($H.num$), the starting offset of the section header table ($H.ofs$) and the size of each section header ($H.sz$). It ends with a section header table, containing a number of fixed-size section headers (line 4).

Each section header contains a starting offset ($SH(i).ofs$) and the size of the corresponding section ($SH(i).sz$). Finally, sections are specified according to their types, offsets, and sizes. Notice that the switch term is used here to decide which nonterminal to use for the specific section according to its section type in the section header.

The detail of the ELF header $H$ is shown starting at line 5 of Figure 9b, which is similar to a C struct, containing a magic number and a list of integer fields. The section header $SH$ is specified in the same way, as shown starting at line 11 of Figure 9b.

For sections, the ELF format has some pre-defined section types. For simplicity, we show only one of them: the dynamic section at line 4 of Figure 9b. If the type of a section ($SH(i).type$) is 6, it is parsed with *DynSec*; otherwise, it is parsed by *OtherSec*, which accepts any input as raw bytes. For a single dynamic section, *DynSec* is a list of *DynSecEntry* as shown at line 15, which is a fixed-size structure and has a similar pattern to *SH*. We omit its detail.

To summarize, the ELF format involves a large degree of random access. Its parser needs to jump from the header to the section header table, then from the section headers inside the table to the corresponding sections. As we have shown, such behaviors can be well captured by IPGs using intervals, attributes, and arrays.

## 4.2 GIF

The Graphics Interchange Format (GIF) is widely used to store bitmap images because of its animation support. We use GIF as an example to show how to use IPGs to specify chunk-based formats. To save space, we do not show the full specification but a subset; the rest is similar.

*Top-level Rule.* GIF starts with a string telling its version to be "GIF89a" or "GIF97a", followed by a block called the Logical Screen Descriptor (LSD), which contains global information including the global color table. After LSD, there is a series of blocks of different types. The number of blocks is unknown until all of them are parsed. Finally, a magic number marks the end of the file. Therefore, the global structure of GIF can be described as follows:

$$GIF \longrightarrow Header[6]\ LSD\ Blocks\ Trailer$$

*Logical Screen Descriptor.* The LSD contains some fixed-size numbers at the beginning and a color table optionally after them. To simplify the presentation, we show only the flag that indicates the existence of a global color table. If the flag is set, the switch term accepts a global color table; otherwise, the empty string is expected.

$$\begin{aligned} LSD \longrightarrow\ & num[1] \\ & \{flag{=}num.val \gg 8\} \\ & \{size{=}3 * (2 \ll (flag \& 7))\} \\ & \mathtt{switch}(flag = 1 : GlobalColorTable[size]\ /\ Empty[0,1]) \end{aligned}$$

*Block List.* The block list is defined recursively by the following rule.

$$Blocks \longrightarrow Block\ Blocks\ /\ Block$$

The first alternative accepts a *Block* with other *Blocks* that start right after the previous one. The recursion ends when it reaches the end of the file, where *Block* would fail, leading to the failure of both alternatives. Then the parser backtracks to the last *Blocks* and tries its second alternative, which succeeds and ends the recursion. Similar specifications occur for other chunk-based formats such as PNG and JPEG. The IPGs of different types of blocks are similar to LSD; we omit their discussion.

## 4.3 PDF

PDF is a more complicated format. Our IPG grammar for PDF does not support full PDF parsing but focuses on how some interesting features in PDF are supported. As a result, the parser generated from our IPG PDF grammar can parse simple PDF files. Two major features of PDF are not implemented: incremental updates and PDF linearization; we believe that they can also be implemented in the IPG framework since they use variants of the random access pattern. There are also semantic properties in the PDF specification that go beyond syntax (e.g., the PDF page tree inheritance property [Adobe 2008]), which cannot be captured by IPGs.

In general, a PDF file contains many PDF objects and an offset table that lists pointers to those PDF objects. The parser of a PDF file starts at the end of the file to get an offset number, which is the starting offset of the offset table. Then following the offsets in the table, the parser can reach all PDF objects.

*Backward Parsing.* PDF has two special parsing patterns. The first one is backward parsing. To parse the offset number at the end of a PDF file, the parser needs to scan backward from the end of the file because the length of the offset is unknown. That is, only where the offset ends are known but not where it starts. One can achieve a backward version of *Num* by the following IPG, which shows our grammar mechanism's flexibility.

$$\begin{aligned} bNum &\rightarrow bNum[\mathbf{EOI}-1]\ Digit[1] \\ &\quad \{v{=}bNum.v * 10 + Digit.v\} \\ &\ \ /\ Digit[\mathbf{EOI}-1, \mathbf{EOI}]\ \{v{=}Digit.v\} \\ Digit &\rightarrow \texttt{"0"}\ \{v{=}0\}\ /\ \texttt{"1"}\ \{v{=}1\}\ \ldots\ /\ \texttt{"9"}\ \{v{=}9\} \end{aligned}$$

*Two-pass Parsing.* The second special pattern is two-pass parsing. Simply speaking, sometimes the length of a PDF object is unknown and stored in another object's header. Therefore, the parser needs to first scan all object headers to get length fields, and then parse all objects again with all lengths known. This pattern can be specified in IPGs because intervals can overlap with each other to let the parser parse the same area more than once.

$$\begin{aligned} &S \rightarrow H[8] \\ &\quad \textbf{for}\ i{=}0\ \textbf{to}\ H.num\ \textbf{do}\ SH[H.ofs + 8 * i, H.ofs + 8 * (i+1)] \\ &\quad \textbf{for}\ i{=}0\ \textbf{to}\ H.num\ \textbf{do}\ OH[SH(i).ofs, SH(i).ofs + 8] \\ &\quad \textbf{for}\ i{=}0\ \textbf{to}\ H.num\ \textbf{do}\ Obj[SH(i).ofs, SH(i).ofs + \exists j.OH(j).link = i\ ?\ OH(j).len : -1] \end{aligned}$$

In this example, the parser first gets the starting offsets of all objects as $SH(i).ofs$ and parses fixed-size object header *OH*. Then in the interval of *Obj*, we use an existential to find the object header $OH(j)$ that satisfies $OH(j).link = i$, which stores the length of $Obj(i)$ as $OH(j).len$. Finally, the parser can complete the parsing of *Obj* after knowing their lengths. This example shows that IPGs allow the parser to solve this kind of data dependency across objects with multiple-pass parsing.

# 5 TERMINATION CHECKING

One advantage of using intervals in our design is that it enables termination checking, which would be difficult in a design where the users were allowed to change the current parsing position arbitrarily. According to IPG parsing semantics, if an IPG parser does not terminate, intuitively there must be a nonterminal whose rule can recursively generate itself; additionally, the intervals must be nondecreasing alongside the generation process. For example,

$$A \rightarrow B[0, \mathbf{EOI}]\ /\ s[0,1];\ B \rightarrow A[0, \mathbf{EOI}]\ /\ s[0,1]$$

obviously does not terminate since the parser iterates between *A* and *B* with the same interval.

Based on this intuition, we design a static termination checking algorithm. If the input IPG contains blackbox parsers, we assume that those blackbox parsers always terminate; their termination checking is delegated to programmers. Then the basic algorithm checks if intervals decrease on all possible paths (with a slight extension, discussed later, to make it less conservative). At a high level, the basic algorithm takes an IPG and conducts the following steps: (1) build a nonterminal dependency graph, (2) enumerate all elementary cycles (defined below) in the graph, and (3) perform symbolic checking on each elementary cycle to ensure termination when IPG parsing follows the cycle of nonterminals during parsing. We next explain the details of these steps.

First, for an input IPG, the algorithm builds a nonterminal dependency graph. It is a directed graph $(V, E)$, where $V$ is the set of nonterminals in the grammar and $E$ contains an edge from $A$ to $B$ labeled with symbolic intervals $[e_l, e_r]$, if $B[e_l, e_r]$ appears in the rule of $A$. Note that there may be multiple edges between two nonterminals.

Second, the algorithm enumerates all elementary cycles in the nonterminal dependency graph. An *elementary cycle* is of the form $A_0 \xrightarrow{(e_{l_0}, e_{r_0})} A_1 \xrightarrow{(e_{l_1}, e_{r_1})} \dots \xrightarrow{(e_{l_{n-1}}, e_{r_{n-1}})} A_n \xrightarrow{(e_{l_n}, e_{r_n})} A_0$, where $A_0 \neq A_1 \neq \dots \neq A_n$. All elementary cycles can be enumerated via depth-first search, or through a more efficient search algorithm [Johnson 1975].

Finally, for each elementary cycle of the above form, the algorithm invokes an SMT solver (we use Z3 [De Moura and Bjørner 2008]) to check if intervals on the cycle always decrease. Note that intervals larger than $[0, EOI]$ such as $[0, EOI + 1]$ are invalid and make the parser stop immediately. Therefore, a non-decreasing cycle must keep looping on the same interval $[0, EOI]$. As a result, we need to check if the following formula is satisfiable

$$(e_{l_0} = 0) \ \wedge \ (e_{r_0} = \mathbf{EOI}) \ \wedge \ \dots \ \wedge \ (e_{l_n} = 0) \ \wedge \ (e_{r_n} = \mathbf{EOI}).$$

If it is unsatisfiable, the intervals decrease when IPG parsing follows the nonterminals in the cycle and the cycle passes the test. If every elementary cycle passes this test, the termination checking algorithm deems that IPG parsing according to the input grammar always terminates; otherwise, it thinks IPG parsing may not terminate. We prove the following theorem and leave its proof to the appendix in the Arxiv version [**?**].

Theorem 5.1. *If static termination checking succeeds for an IPG with a start nonterminal $A_0$, for any input string $s$, parsing with "$s \vdash A_0 \Downarrow R$" terminates with some result $R$.*

The basic termination checking algorithm is conservative. We describe a slight extension for the special **end** attribute. Recall that in the GIF format we have a rule

$$Blocks \longrightarrow Block[0, \mathbf{EOI}] \ Blocks[Block.\mathbf{end}, \mathbf{EOI}].$$

If $Block.\mathbf{end}$ is treated as a normal attribute, then this rule cannot pass the termination checking because $Block.\mathbf{end}$ can be 0. In our extended termination checking, if the rule of a nonterminal $A$ consumes at least one terminal (which can be checked by a syntactic check), we add a conjunctive clause $A.\mathbf{end} > 0$ to the formula sent to the SMT solver. This is the case for $Block.\mathbf{end}$ in the above example, which ensures the termination of interval parsing for GIF.

## 6 RELATED WORK

In this section, we first discuss the origins of some of the IPG's ideas, then compare existing efforts on specifying data dependence and parsing file formats with the IPG.

### 6.1 Formal Grammars

Generating parsers from declarative grammars has been researched for decades and the most widely-used grammar is Context-Free Grammars (CFGs). Some techniques focus on accepting

```
# FAILURE: We can't go back to parse secs!
simplified_elf = { o=num } { offset=string2int(o) } skip(offset) sec_hdrs secs
skip(n) = ([n > 0] CHAR8 {n = n - 1})* [n = 0]
secs = ...
sec_hdrs = ...
num = digit digit*
digit = 0 | 1
```

Fig. 10. Attempt to use YAKKER to specify ELF.

any CFGs, including Earley parsing, GLL, and GLR, among others. There are also techniques that restrict CFGs in various ways to cope with issues such as ambiguity, including LL(k), LL(*), LR(k), SLR, and LALR [Aho et al. 1986]. However, CFGs cannot capture dependencies required in file format specifications, such as the random access pattern and the type-length-value pattern.

Main design choices of our IPG framework include recursive descent parsing, a biased-choice operator, attributes, and intervals. We next briefly relate these design choices to prior work. First, we note that traditional recursive descent parsing does not allow left recursive grammars. In contrast, our IPGs allow left recursion as long as there are no cycles with non-decreasing intervals (see section 5 about termination checking). Second, our IPGs rely on a biased-choice operator to resolve ambiguity, similar to Parsing-Expression Grammars (PEGs [Ford 2004b]). PEGs have difficulty in specifying file formats, same as CFGs. PEGs can specify the reverse length-value pattern in a complicated way [Lucas et al. 2021], but whether they can specify the length-value pattern remains unknown, let alone the random access pattern.

Third, IPGs rely on attributes for propagating dependencies. Attribute grammars were originally proposed by Knuth [1968] for modeling programming language semantics. An attribute grammar specifies how attributes are computed for terminals/nonterminals based on attributes of neighboring nodes in a parse tree, so that information can flow across the parse tree. Attributes provide a mechanism for context sensitivity in a CFG. IPGs support *synthesized attributes*; the value of a synthesized attribute of a node in a parse tree is computed from the values of the attributes of the node's children in the tree (i.e, through bottom-up propagation). Attributes in conventional attribute grammars do not affect the parsing process as attributes can in theory be computed after a parse tree is constructed. By using attributes in intervals, our IPGs' attributes can directly affect the parsing process, which is required by many file formats.

### 6.2 Comparison with Other Data-Description Languages

*Data-dependent grammars* can capture dependencies in input data. For example, PADS [Fisher and Gruber 2005; Fisher et al. 2006] is a dependently typed data-description language that allows later parsing to depend on values produced earlier. Yakker [Jim et al. 2010] formalized the translation of data-dependent grammars to data-dependent automata, with an Earley-style parsing algorithm. Some format description languages like Parsley [Mundkur et al. 2020] and EverParse [Ramananandro et al. 2019] allow limited forms of data dependencies (e.g., the type-length-value pattern). Finally, attribute grammars by Underwood [2012] for binary file formats use attributes in predicates to indicate the number of repetitions, similar to Yakker's dependent fields.

However, data-dependent grammars cannot fully capture patterns in file formats. They are fundamentally limited by their assumption of the left-to-right parsing order. To be specific, consider the attempt of using Yakker [Jim et al. 2010] to specify a simplified ELF format as shown in Figure 10. The first line is the top-level rule that first parses a 0-1 string `o` and converts it to a number `offset`.

Then skip(offset) is used to skip that length of the input, followed by sec_hdrs, whose detail is omitted. However, there is no way to specify secs because that part of the input has already been skipped and the parsing in Yakker cannot go backward. Another example is the backward parsing in the PDF format in section 4, which is also out of the ability of data-dependent grammars.

Another line of work is the declarative data formats tools such as Kaitai Struct [Kaitai 2015], Nail [Bangert and Zeldovich 2014], DataScript [Back 2002] and FlexT [Hmelnov and Mikhailov 2019], designed to describe binary formats. To accommodate the random access pattern, they provide labels for each structure to indicate its starting offset in the file. This is in spirit similar to the use of a seek operator, which allows the parser to move the parsing position to an arbitrary user-specified position.

```
seq:
  - id: name
    type: subparser
types:
  subparser:
    seq:
      - id: offset
        type: u1
    instances:
      jump:
        io: _root._io
        pos: offset
        type: subparser
```

(a) Seeking in Kaitai Struct.

$$S \rightarrow num[0,1]\ S[num.val, \mathbf{EOI}]$$
$$num \rightarrow \{v{=}btoi[0,\mathbf{EOI}]\}$$

(b) An IPG equivalent to Figure 11a.

```
seq:
  - id: name
    type: epsilon
    repeat: eos
types:
  epsilon: {}
```

(c) Repeating epsilon in Kaitai Struct.

$$S \rightarrow \texttt{""}[0,0]\ S[0,\mathbf{EOI}]$$

(d) Repeating epsilon in IPGs.

Fig. 11. Nontermination examples in Kaitai Struct and their counterparts in IPGs.

Although the seek operator enhances expressiveness, it falls short in other aspects because of its imperative nature (the seek operator is similar to a goto in a program). We use Kaitai Struct as an example. First, it would be difficult to check termination if the parsing position can be moved arbitrarily through a seek operator. The Kaitai Struct snippet in Figure 11a generates a parser that calls the subparser, which is defined to consume an unsigned 1-byte integer u1 with the field name offset. And there is another construct called jump in the definition of subparser. It redirects the current input stream to the global input stream by io: _root._io, then moves the parsing pointer to offset, and finally calls subparser again. If the given input is "0", then the generated parser will keep consuming the first byte then moving back to it until the stack overflows. For comparison, we list the equivalent IPG in Figure 11b. where btoi is the specialized function for integer parsing as explained in section 7. It is easy to see its nontermination since $num.val$ can be 0.

Another nontermination example is in Figure 11c, which specifies a parser that repeats the subparser epsilon until the eos (end of the stream). However, the subparser epsilon consumes nothing ({}); so it keeps parsing an empty string until running out of memory. On the contrary, using intervals as the IPG shown in Figure 11d, one can immediately tell its nontermination since $[0,\mathbf{EOI}]$ is always the same interval.

Furthermore, the seek operator breaks modularity and therefore reduces the readability of specification. Take ELF in section 4 as an example. To achieve random accessing from an ELF header to its sections, Kaitai uses the `jump` pattern in Figure 11a, It essentially uses gotos, which breaks modularity and makes it hard to reason about the specification. Instead, the IPG encapsulates each component in its own rule while expressing their dependency in the form of attribute referencing between different nonterminals in the same alternative.

We believe that IPGs reduce programmers' cognitive workload when reading and writing grammars, in comparison with imperative-style tools such as Kaitai. For reading, since an IPG expresses a tree structure where dependencies occur only between siblings, it enables readers to understand the whole grammar in a top-down fashion. In contrast, in Kaitai it is often the case that readers need to read through the whole specification because of its imperative nature. The reasoning process during reading is like following imperative instructions step by step in mind; before going through the entire process it is difficult to understand individual components. For writing, the experience of writing IPGs (functional style) and Kaitai specifications (imperative style) varies from person to person. However, when converting an informal specification into a formal one, the programmer typically goes through the process by trial and error, which means that he/she needs to read through all that has been written to check if it matches the informal specification. In this case, we think the advantage of IPGs being easier to read would be beneficial.

## 7 IMPLEMENTATION AND EVALUATION

We implemented in OCaml an IPG parser generator (publicly available at [Zhang et al. 2023]), which generates C++ recursive descent parsers in a standard way [Aho et al. 1986]. Every nonterminal is translated to a C++ function, which checks terminal strings and calls functions for other nonterminals according to its rule. We specialized the parser for integers as a special function `btoi` to replace the inefficient *Int* implementation that we have shown in previous examples, since it is frequently called in practice.

We already compared IPGs with other tools in terms of expressiveness in section 4 and section 6. For additional comparison in terms of the specification size and the performance of the generated parsers, we compared IPGs with Kaitai Struct [Kaitai 2015] and Nail [Bangert and Zeldovich 2014]. Kaitai Struct is a popular tool for parsing file formats and has been used to parse over 150 formats. Nail mainly focuses on network packet formats and generates efficient parsers. We compared IPGs against Nail only on network packet formats since Nail does not support general file formats. The parsers generated by Kaitai Struct and Nail are in C++ and C, respectively. We do not include other data description languages such as PADS and P4 during comparison since our work focuses on file format parsing, which is generally not supported by these tools as we have shown in section 6. Furthermore, we examined the effectiveness of implicit intervals by checking how many interval annotations can be inferred and omitted.

During evaluation, we used IPGs to specify a number of popular file formats (ZIP, GIF, PE[1], ELF, and part of PDF) and network packet formats (IPv4+UDP and DNS). Notice that we specified only the syntax and the parsing-related semantics of these formats but not other semantics like file validation. For example, checksums are commonly used in network packet formats for data integrity. Although they can be specified in IPGs through predicates, we do not include them in the specifications of network packet formats because they do not affect the parsing process and can be checked through a separate validation pass. The IPG grammars of all these formats passed termination checking, with less than 20ms for termination checking because these grammars had no more than five elementary cycles.

*Specification Sizes.* We show lines of code for formats under evaluation by Kaitai Struct and IPGs in Table 1. While this metric is not perfect for evaluating programmer effort, to the best of our knowledge, there are no alternative low-cost metrics. In general, having fewer lines of code at least means that the specification is more compact.

[1] PE stands for Portable Executable, the standard binary format on Windows.

Table 1. Lines of format specifications.

| | ZIP | GIF | PE | ELF | PDF | IPv4+UDP | DNS |
|---|---|---|---|---|---|---|---|
| IPG | 102 | 61 | 109 | 96 | 108 | 22 | 34 |
| Kaitai | 256 | 163 | 223 | 244 | N/A | 69 | 105 |
| Nail | N/A | N/A | N/A | N/A | N/A | 26+29 | 39+60 |

Table 2. Number of Intervals and Implicit Intervals in IPG Specifications.

| | ZIP | GIF | PE | ELF | PDF | IPv4+UDP | DNS |
|---|---|---|---|---|---|---|---|
| Intervals | 87 | 55 | 97 | 82 | 241 | 17 | 28 |
| Implicit Intervals | 14+55 | 20+26 | 4+81 | 5+48 | 116+83 | 1+14 | 4+14 |

Nail has only limited features and relies on an extension written in C to support patterns like the type-length-value pattern. So we show Nail's line of code in the form of $a + b$, where $a$ is the line of code in Nail and $b$ is the line of code in C. Further, we compared with Nail only on network packet formats since it would require a large amount of hand-written C code to support complicated file formats, which deviates from our intention of this comparison. Finally, Kaitai Struct did not come with a PDF specification; so we did not compare with it on the PDF format.

Kaitai Struct has more lines of specifications than IPGs mainly for two reasons: (1) A Kaitai Struct specification is similar to the declaration of C structs; when a format has the random access pattern or recursion such as labels in DNS packets, Kaitai Struct needs lengthy imperative annotations like the example in Figure 11a; (2) Kaitai Struct uses line breaks to separate fields (similar to Python); so the total number of lines becomes larger.

We use the ZIP decompressor as a concrete example. Linux's `unzip` implementation has its main routines in a file that has 1,600 lines of C code, which parses the metadata and calls decompression methods in other files. By using IPGs, we achieved the same functionality in 102 lines of IPG code, calling a blackbox parser that calls `zlib` for decompression in about 70 lines of C code as mentioned in section 3.4.

*Effectiveness of Implicit Intervals.* We show the effectiveness of implicit intervals in Table 2. In the first line, we show the total number of intervals used in an IPG grammar; the second line shows how many intervals are eliminated by implicit intervals in the form of $a + b$, where $a$ is the number of intervals that are fully eliminated, and $b$ is the number of intervals that include only a length. Overall, 27.0% of intervals can be fully eliminated and 52.9% of intervals need only the length specification. This shows implicit intervals' effectiveness in reducing developer burdens.

In practice, we find that implicit intervals are mainly useful in two cases. The first case is the CFG-like behavior where we have a sequence of nonterminals or terminals that appear one after another. With implicit intervals, we do not need to write **end** repeatedly. For the second case, many formats start with magic numbers, which are terminal strings at the beginning of alternatives. Their intervals can be inferred by auto completion.

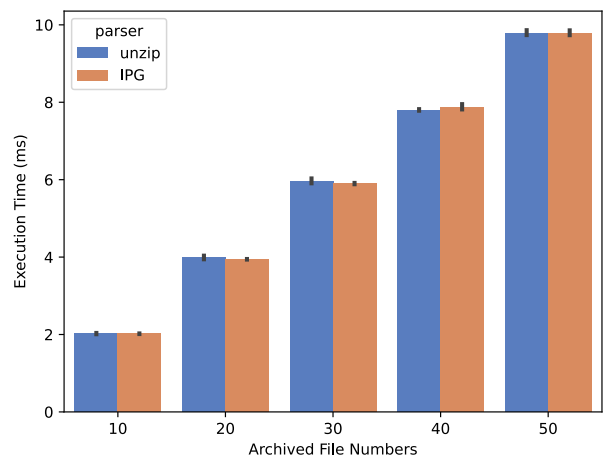

(a) unzip

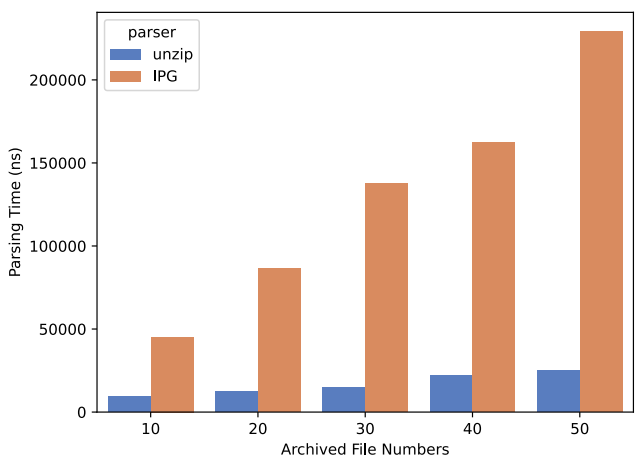

(b) parsing time of unzip

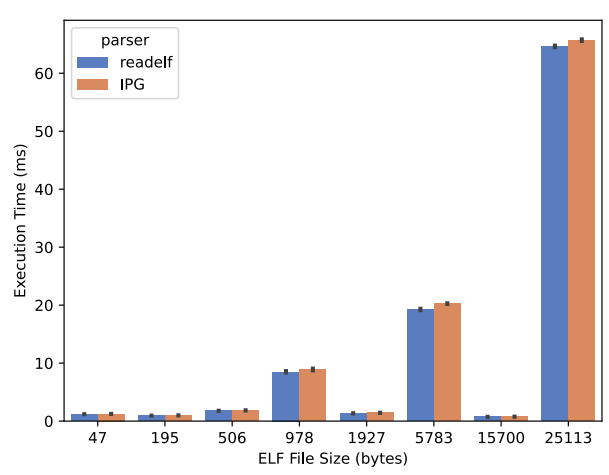

(c) readelf -h -S -dyn-syms

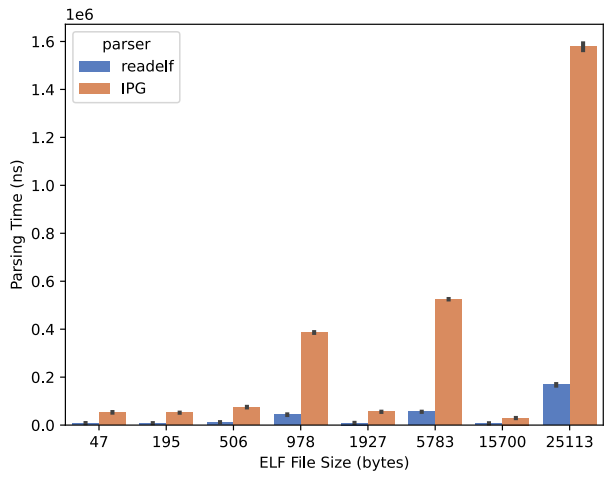

(d) parsing time of readelf

Fig. 12. Performance comparison with hand-written parsers.

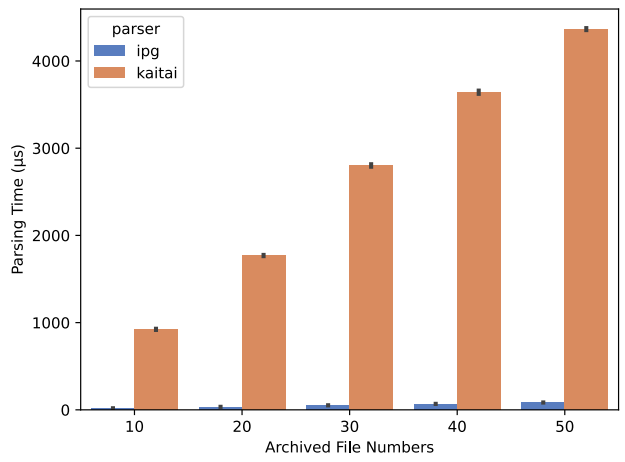

(a) ZIP

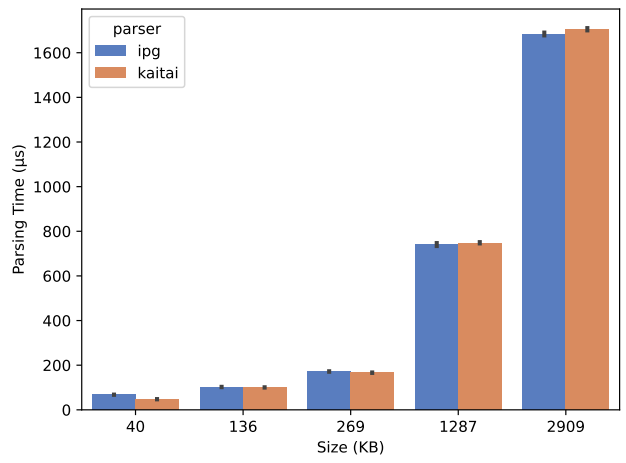

(b) GIF

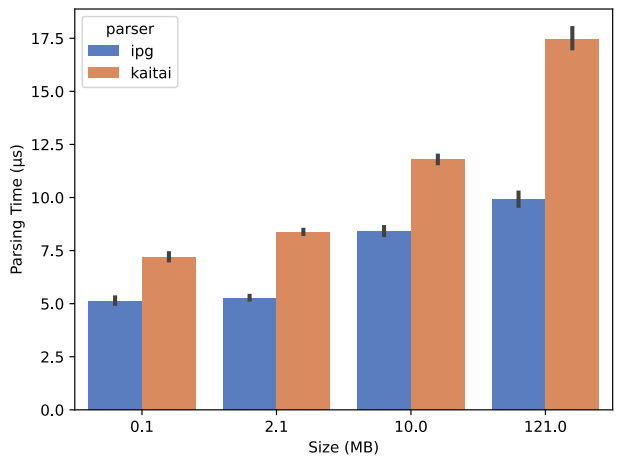

(c) PE

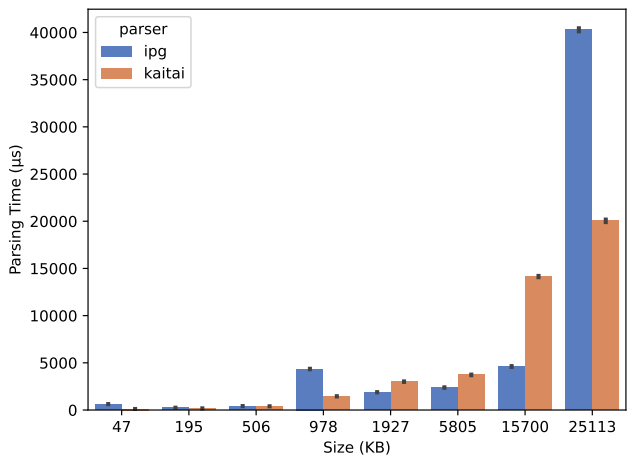

(d) ELF

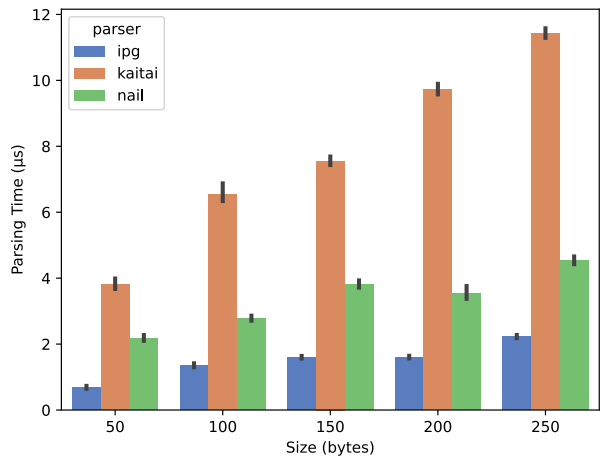

(e) DNS

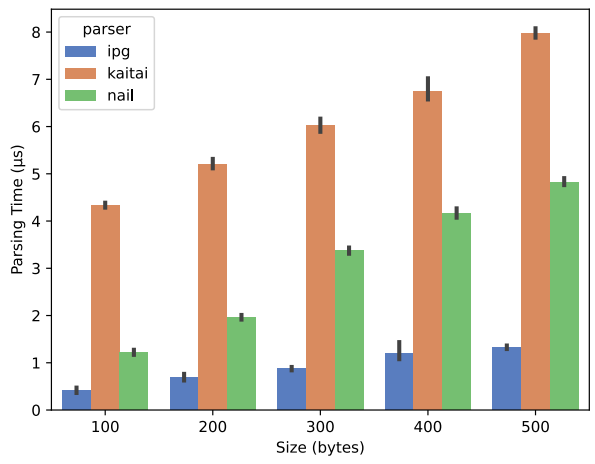

(f) IPv4+UDP

Fig. 13. Parsing Time for Different Formats.

*Performance.* To evaluate the performance of generated parsers against handwritten parsers, we modified two open-source Linux utilities, `readelf` and `unzip`, by replacing their parsing components with parsers generated from our IPG grammars. The end-to-end running time and parsing time results are in Figure 12. In the `unzip` case, the end-to-end time includes time spent on parsing the file, decompression and writing files. In the `readelf` case, it includes parsing the file, validating data, and printing results. To isolate the parsing time of `readelf` and `unzip`, we manually identified the components that belong to the parsing, measured their runtime times, and summed them up. As handwritten tools such as `readelf` and `unzip` tightly mix parsing and the following processing steps, our isolation of parsing time might be an underapproximation. The results show that the hand-written parsers are much faster but using IPGs does not downgrade the end-to-end performance because the parsing time is a small part of these two applications. The parsing time performance gap might be due to the fact that the hand-written parsers directly map file data to C struct fields, allowing them to benefit from sequential disk I/O. We leave this optimization for IPG parsers to future work.

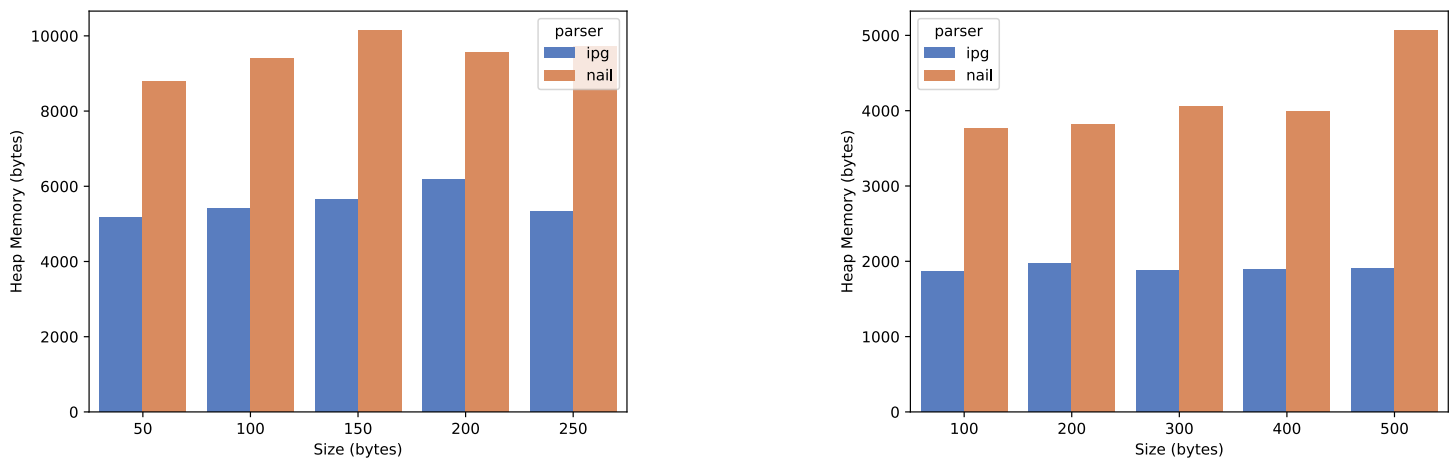

(a) DNS (b) IPv4+UDP

Fig. 14. Heap Memory Consumption for Packet Parsing

We also evaluated the performance of parsers generated from IPGs with parsers generated from Kaitai Struct. We developed parsers for ELF, PE, ZIP, and GIF through parser generation via IPGs and Kaitai Struct specifications. For IPv4+UDP and DNS, we also included parser generation via Nail. We evaluated all parsers on various sizes of sample files (in the ZIP case, different numbers of compressed files). For PE and ELF, samples are executable files for real applications on Windows and Linux respectively. GIF samples are pictures collected from the Internet. ZIP samples archive different numbers of copies of the same file. IPv4+UDP and DNS samples are real network packets captured by a packet analyzer on our machine. In all these cases, we measured only the parsing time, which is the running time of the parse function provided by these tools. Also, to exclude disk I/O time, we read the entire file into memory before the timer starts. The average parsing running times for 1000 runs are shown in Figure 13 as well as the variance. The test machine has an Intel i7-9700 CPU and 32 GB memory.

The performance result is shown in Figure 13. For network packets, Nail implements arena-based memory management to avoid performance impact from calling `malloc`. In both network packet formats, IPGs achieve better running time after adopting the same mechanism.

For other file formats, IPGs have similar performance as Kaitai Struct on GIF and PE cases (shown in Figure 13b and Figure 13c respectively). For ZIP, as Figure 13a shows, Kaitai Struct performs much worse than IPGs because its implementation consumes the archived file data to move the input position to the next section, while IPGs generate a zero-copy parser that just skips archived file data. To the best of our knowledge, there is no option to let Kaitai Struct generate zero-copy parsers. On the other hand, for decompression after parsing, using the IPG-generated parser we would still need to read the compressed data into memory, while using the Kaitai Struct parser the compressed data would already be resident in memory; therefore the parsing performance difference between IPGs and Kaitai Struct is only part of the story. For ELF parsing, the IPG parser has similar performance to the Kaitai Struct parser for small and medium-sized files. For the largest test file (over 25MB), the IPG parser performs much worse because parsing symbol names requires deep recursion in the IPG parser, which could be eliminated by introducing the Kleene-star operator into IPGs. We leave this as a future optimization.

To validate the correctness of IPG parsers, in the `unzip` case, the decompressed files were manually checked. In the `readelf` case, the output of the modified `readelf` with our IPG parser was compared with the output of the original `readelf`. In other formats' cases, the output parse tree was compared with Kaitai Struct's.

For packet parsing, using Valgrind we also measured the amount of heap memory consumption in comparison with Nail. The results are shown in Figure 14; they show that IPG parsers consume less memory than Nail parsers for both dns and ipv4+UDP.

## 8 CONCLUSION AND FUTURE WORK

Interval Parsing Grammars provide a declarative and hierarchical way of expressing the syntax and parsing-related semantics of file formats. With attributes and intervals, IPGs allow the specification of data dependence as well as the dependence between control and data. Moreover, parser termination checking becomes possible. To further utilize the idea of intervals, an interval-based, monadic parser combinator library is proposed. Using these techniques, IPGs can support many file formats.

We believe that the declarative nature of IPGs enables its metatheory to be fully mechanized in an interactive theorem prover such as Coq so that we can get verified IPG parsers with correctness and termination guarantees. For correctness, we can prove that the generated parser for an IPG recognizes the same set of strings and builds the same parse trees as a denotational semantics of the IPG dictates. The functional nature of IPGs makes defining the denotational semantics and proving correctness much easier. Further, although this paper focuses on the parsing of file formats, to achieve the goal of formalized format specifications, there are more semantic constraints besides the parsing behaviors discussed in this paper, ranging from simple value restrictions to more complicated constraints on data structures generated by parsing. We can combine IPGs with a recently proposed declarative language for checking semantic constraints [Kumar et al. 2023]. Another interesting direction is to automatically generate a file generator from the given grammar, which turns the internal representation produced by the parser back to a file. This idea has been explored by Nail [Bangert and Zeldovich 2014]. To support this feature, IPGs can be extended with reverse attribute computation that specifies how an attribute is computed in the reverse direction of parsing. With attributes computed, intervals are known and writing terminals back to the file according to their intervals is straightforward. Finally, while our parser generator generates only recursive descent parsers, it would be interesting to explore the support of stream parsers. We can first have an analysis that determines if it is possible to generate a stream parser from an IPG: within each production rule, it checks if the attribute dependency is only from left to right. After this analysis, a stream parser can be generated to parse in a bottom-up way. We plan to explore these directions in future work.

## ACKNOWLEDGMENTS

We thank anonymous reviewers and our shepherd, Francois Gauthier, for in-depth comments on earlier versions of the paper. Also, we would like to thank the artifact evaluation committee for numerous comments that greatly improved the artifact. This research is supported by DARPA under Grant No. HR0011-19-C-0073 and NSF under Grant No. FMiTF-1918396.

## ARTIFACTS

The code and evaluation results are available on Zenodo [Zhang et al. 2023]. To reproduce the result or reuse the implementation, please follow the instructions in its README file.

$\boxed{s, A, E, \overrightarrow{Tr} \vdash tm \Downarrow E, R_{\text{tm}}}$

$$\frac{\begin{array}{c}\sigma(E, \overrightarrow{Tr}, e_1) = i \qquad \sigma(E, \overrightarrow{Tr}, e_2) = j \\ \forall i \leq k < j, \\ \sigma(E[id \mapsto k], \overrightarrow{Tr}, e_l) = l \qquad \sigma(E[id \mapsto k], \overrightarrow{Tr}, e_r) = r \qquad s[l, r] \vdash B \Downarrow \text{Node}_\text{k}(B, E_k, \overrightarrow{Tr_k})\end{array}}{\begin{array}{c}s, A, E, \overrightarrow{Tr} \vdash \texttt{for } id\text{=}e_1 \texttt{ to } e_2 \texttt{ do } B[e_l, e_r] \Downarrow \\ \text{updStartEnd}(\ldots(\text{updStartEnd}(E, l + E_i[\textbf{start}], l + E_i[\textbf{end}], E_i[\textbf{end}] \neq 0), \ldots), \\ l + E_{j-1}[\textbf{start}], l + E_{j-1}[\textbf{end}], E_{j-1}[\textbf{end}] \neq 0), \\ \text{Array}(\text{Node}(B, E_i[\textbf{start} \mapsto l + E_i[\textbf{start}]), \textbf{end} \mapsto l + E_i[\textbf{end}])], \overrightarrow{Tr_i}) \cdot \ldots \cdot \\ \text{Node}(B, E_{j-1}[\textbf{start} \mapsto l + E_{j-1}[\textbf{start}]), \textbf{end} \mapsto l + E_{j-1}[\textbf{end}])], \overrightarrow{Tr_{j-1}}))\end{array}} \text{(T-ArraySucc)}$$

$$\frac{\begin{array}{c}\sigma(E, \overrightarrow{Tr}, e_1) = i \qquad \sigma(E, \overrightarrow{Tr}, e_2) = j \\ \exists k,\ i \leq k < j \\ \sigma(E[id \mapsto k], \overrightarrow{Tr}, e_l) = l \qquad \sigma(E[id \mapsto k], \overrightarrow{Tr}, e_r) = r \qquad \neg(0 \leq l < r \leq |s|) \ \vee \ s[l, r] \vdash B \Downarrow \text{Fail}\end{array}}{s, A, E, \overrightarrow{Tr} \vdash \texttt{for } id\text{=}e_1 \texttt{ to } e_2 \texttt{ do } B[e_l, e_r] \Downarrow E, \text{Fail}} \text{(T-ArrayFail)}$$

Fig. 15. Full IPG parsing semantics.

# A APPENDIX

Theorem A.1. *If static termination checking succeeds for an IPG with start nonterminal $A_0$, for any input string $s$, parsing with "$s \vdash A_0 \Downarrow R$" terminates with some result $R$.*

Proof. We define parse$(s, A) = R$ if "$s \vdash A \Downarrow R$" according to the rules in Figure 8. Inspecting those rules, we know parse$(s, A)$ performs a call to parse$(s', B)$ only when $B[e_l, e_r]$ appears in the IPG grammar rule of $A$, in which case by construction the edge with label $(e_l, e_r)$ from $A$ to $B$ appears in the dependency graph. Therefore, any call to parse$(-, -)$ is faithfully modeled in the dependency graph.

Starting from nonterminal $A_0$, suppose a parse$(-, -)$ call chain is $A_1$, $A_2$, etc. Suppose there are $n$ nonterminals in the grammar. Then when the call chain depth is at most $n$, we encounter the first elementary cycle (if there is any). Suppose the cycle is $A_i \to \ldots \to A_j \to A_i$. Because of static termination checking, we know when $A_i$ is visited the second time, the input string size is at least one less. With the same argument, from the second visit to $A_i$, with the call chain depth at most $n$, we encounter another elementary cycle (if there is any), which also decreases the input string size by at least one. Therefore, the depth of any parse$(-, -)$ call chain during parsing is bounded by $n * |s|$. Coupled with the fact that parsing a nonterminal results in a finite number of direct calls to parse$(-, -)$, we can derive that IPG parsing terminates for $s$. □

## A.1 Full Parsing Semantics

Figure 15 is the full formalism of IPG's parsing semantics. Only those rules that are different from the short version are listed.

## A.2 Interval Parser Combinators

Parser combinators provide another popular way of implementing parsers. A parser combinator library provides some basic parsers (e.g., a parser that parses a single character) and some combinators (e.g., a choice operator), and the user can construct a complete parser using these

```
type state = int * int * int
type 'a parser =
  string -> state -> ('a * state) option
let return (v:'a) : 'a parser =
  fun inp s -> Some (v,s)
let bind (m: 'a parser) (f: 'a -> 'b parser)
    : 'b parser = fun inp s ->
    match m inp s with
    | Some (v,s1) -> f v inp s1
    | None -> None
let (>>=) = bind

let getInterval: (int*int) parser =
  fun inp (l,r,c) -> Some ((l,r),(l,r,c))
let setInterval (l:int) (r:int): unit parser
  = fun inp _ ->
      if l < r then Some ((),(l,r,l))
      else None
let getPos: int parser = fun _ (l,r,c)
  -> Some (c,(l,r,c))
let setPos (c:int) : unit parser =
  fun inp (l,r,_) -> Some ((), (l,r,c))
```

Fig. 16. Monadic Parser Combinators.

building blocks. Compared with the parser generator approach, it is less declarative since how parser combinators are combined is expressed in a general-purpose programming language. On the other hand, a file format can involve arbitrarily complex parsing behaviors. Parser combinators enable users to write expressive parsers to cover these behaviors in a general-purpose programming language.

We present an interval-based parser-combinator library in Figure 16. It is written in OCaml, but the same design can be adopted by other programming languages. As many other parser combinator libraries, this library is implemented in a monadic style [Hutton and Meijer 1998]. A parser of type `'a` takes a string and a state as input and, if parsing succeeds, produces a value of type `'a` and an output state; if parsing fails, `None` is returned. The internal state of the monad is a triple `(l,r,c)`, where `l` and `r` are the left and right endpoints of the interval assigned to the parser and `c` is the current parsing position. The return/bind combinators are similar to traditional parser combinators.

Figure 16 also presents a set of basic combinators for manipulating the internal state of a parser. `getInterval`/`setInterval` gets/sets the interval, i.e., `l` and `r` in the monad state. Similarly, `getPos`/`setPos` gets/sets the current parsing position, i.e., `c` in the monad state. However, these combinators operate directly on the absolute offsets of the global input and do not match IPG's parsing semantics, where an interval uses offsets relative to the current local input. Therefore, these combinators should not be used by users and the library provides other combinators on top of them as the user interface. Some example combinators follow:

```
let eoi =
    getInterval >>= fun (l,r) ->
    return (r-l)
```

```
let localIntervalP p (l,r) =
    getInterval >>= fun (l_g,r_g) ->
    if 0<=l && r<=r_g-l_g then
        setInterval (l_g+l) (l_g+r) $$
        p >>= fun v ->
        setInterval l_g r_g $$
        setPos (l_g+r) $$
        return v
    else fail
let (%) = localIntervalP
```

eoi returns the end-of-input as a relative offset, which is the length of the local input. Combinator "%" runs parser p in a local interval [l,r], where l and r are relative offsets regarding the current local interval; the semantics of this combinator matches the IPG semantics. Internally, the definition of "%" first gets the current interval [l_g,r_g], checks if the new interval is in the range of the current interval, calls parser p on the new interval, recovers the old interval, and finally returns the result of the parser. In the definition, $$ sequences two parsers and ignores the value returned by the first one; its definition is straightforward and omitted. As an example of using eoi and %, "$a[3, EOI]$" in an IPG can be written as "eoi >>= fun ed -> a % (3, ed)".

There are more combinators including biased choice /, sequence $, array arr, etc. Their semantics and implementation are straightforward; so we omit their definitions.

*An Example.* Consider the previous binary number parser in Figure 3, We can write a parser with the same functionality using parser combinators.

```
let intP = fix (fun intp ->
  eoi >>= fun eoi ->
  intp % (0, eoi-1) >>= fun iv ->
  digitP % (eoi-1, eoi) >>= fun dv ->
  return (iv * 2 + dv)
  / (digitP % (0,1) >>= fun dv ->
  return dv))
let digitP = charP '0' % (0, 1)
           / charP '1' % (0, 1)
let charP (ch:char): char parser =
  fun inp (l,r,c) ->
  if (l<=c && c<r && inp.[c]=ch) then
    Some (ch, (l,r,c+1))
  else None
```

This example shows features in IPGs can be mapped to combinators; e.g., attributes are bound with their names using the bind operator >>=; intervals are assigned to subparsers using %. In this way, we have implemented all case studies in section 4 through our parser combinator library.